\documentclass[journal=nalefd,manuscript=article,layout=onecolumn,
maxauthors=10]{achemso}

\usepackage{chemformula} 
\usepackage[T1]{fontenc} 
\usepackage{diagbox} 
\usepackage{changes}

\usepackage{amsmath}
\usepackage{amssymb}
\usepackage{hyperref}
\usepackage{graphicx}
\usepackage{dcolumn}
\usepackage{bm}
\usepackage{color}
\usepackage{lineno}

\author{Ke Zhou}%
\email{zhouke@suda.edu.cn}
\affiliation
{
College of Energy, Soochow University, Suzhou 215006, China
}

\title{
Ion Clustering Regulated by Extreme Nanoconfinement Enables Mechanosensitive Nanochannels
}%


\begin{document}



\newpage

\begin{abstract}

Mechanosensitive ion nanochannels regulate transport by undergoing conformational changes within nanopores. However, achieving precise control over these conformational states remains a major challenge for both artificial soft or solid pores.
Here, we propose an alternative mechanism that modulates the charge carrier density inside nanopores, inspired by transistors in solid-state electronics. 
This strategy leverages a novel phenomenon of confinement-regulated ion clustering in two-dimensional extremely confined nanochannels, revealed by extensive $\mu$s-scale enhanced-sampling molecular simulations based on an \emph{ab initio}–refined force field and nucleation theory.
The resulting \emph{force-ion transistor} enables mechanically gated control of ion transport and provides a conceptual foundation for designing ionic mechanical logic gates. 
Our findings offer new insights into piezochannel mechanosensing and electromechanical coupling in biosystems beyond conformational signaling, opening pathways to integrate artificial ion channels with neuromorphic devices for processing mechanical stimuli.

\end{abstract}

\newpage

\section*{Introduction}

Ion channels are protein structures with nanoscale pores embedded in cell membranes that enable selective ion transport, facilitating multifunctional cellular sensors and the transmission of electrical signals \cite{hille2001_ionch,epand2018}.
These channels play essential roles in response to chemical and mechanical stimuli by regulating the transport of ions, mediating various biological responses such as  nerve impulse transmission, muscle contraction, and maintaining chemical balance in cells \cite{hille2001_ionch,epand2018,syeda2016_cr,murthy2017_nrmcb}.
Mechanically activated ion channels, such as such as PIEZO, TREK
and TRAAK, confer force sensitivity to cells and organisms by tuning the passage of ions across the membrane in response to a mechanical stimulus \cite{murthy2017_nrmcb,wu2017,brohawn2014_pnas,saotome2018_nature}.
Replicating such mechanosensitive functions with artificial nanochannels holds great promise for versatile applications, including bio-sensing, precision drug delivery, responsive diagnostics, neuronal-computer interfaces, and advanced computing \cite{murthy2017_nrmcb,bocquet2020_natmater2,
bocquet2021_science,duan2022_pnas,bocquet2023_science,zhou2020_nanoletter}. 
Beyond technological applications, such artificial analogs also provide a model platform for advancing the fundamental understanding of mechanotransduction in living systems \cite{hille2001_ionch,murthy2017_nrmcb}.

Over the past decades, advances in nanofabrication have reduced the dimensions of fluidic devices down to the unprecedented nanometre- or even {\AA}ngstrom-scale. 
Nanotubes, nanopores, or nanoslits are created by direct synthesis, lithography or self-assembly, which led to the discovery of unexpected phenomena in water and ion transport processes \cite{munoz2021_chemrev,bocquet2023_science,bucquet2022_natmater,
bocquet2019nature,bunch2012_natnano,holt2006_science,
tunuguntla2017_science,zhao2021_nc}.
Notably, artificial slit-like channels with {\AA}ngstrom-level precision have been realized via van der Waals (vdW) assembly of two-dimensional (2D) materials, whose atomically smooth and chemically tunable surfaces offer a uniquely controlled confined environment \cite{geim2017_science,geim2018_science}.
This technical breakthrough allows us to explore ionic processes in extremely confined systems, where water can exhibit layered ordering, or even exist as a monolayer (1L).
Because the dimensions of these channels are comparable to, or even approach, the molecular scale of water and hydrated ions, they exhibit exotic ionic transport behaviors beyond classical expectations. Such phenomena include memristive effects, ionic Coulomb blockade, rectification, and ionic logic operations \cite{bocquet2021_science,bocquet2023_science,bucquet2022_natmater,bocquet2019nature,bocquet2019_natnano,zhou2020_nanoletter,zhao2021_nc}.
Although remarkable progress has been made in designing artificial biomimetic devices over the past decade, the realization of artificial mechanotransductive iontronic systems remains scarce \cite{murthy2017_nrmcb}.

One could envision the development of artificial mechanotransduction iontronic devices that mimic biological channels, which regulate ion passage through structural changes in response to osmotic pressure or membrane tension \cite{murthy2017_nrmcb,wu2017,brohawn2014_pnas,saotome2018_nature}. 
However,the reported modulation of conductance in these devices remains limited, especially for low modulation of conductance \cite{davis2020_nanolett,li2023_jacs}.
Reproducing the precise structure and function of biological channels remains a challenge, highlighting the need for innovative approaches to ion regulation in the design of mechanosensitive nanochannels.

Ions behave similarly to an ideal gas with low ion-ion correlation at low or moderate concentrations in bulk water. However, under nanoconfinement, ion-ion correlation increases, leading to slow structure relaxation and the formation of paired or clustered states \cite{bocquet2019_natnano,
zhou2020_nanoletter,zhao2021_nc,bocquet2021_science}.
This indicates that ionic dynamics, configurations of ions, and ultimately, ion passage can be modulated by confinement \cite{bocquet2019_natnano,
zhou2020_nanoletter}.
We investigate the solvation and dynamics of alkali-metal-halide electrolytes in 2D nanochannels with interlayer spacings ($h$) comparable to the size of water molecules and hydrated ions.
We observe ions spontaneously forming ion pairs and clusters when $h< 0.95$ nm or on 1L water.
$\mu$s-scale enhanced-sampling molecular dynamics (ESMD) with \emph{ab-initio} derived force fields show that ions in these confined layers prefer fully clustered states, while in bulk water, they remain partially hydrated.
The clustering free energy is sensitive to confinement, suggesting that ion density can be regulated by adjusting $h$.
We use these findings to design a mechanosensitive ion channel where ion clustering is controlled by mechanical stimuli, achieving an on-off ratio of approximately $100$.
This concept also enables the creation of ionic mechanical logical gates and provides foundational principles for mechanosensing in biosystems beyond simple conformational changes \cite{epand2018}.

\section{Results and Discussions}

\begin{figure*}[tbhp]
  \centering
  \includegraphics[width=0.90\textwidth]{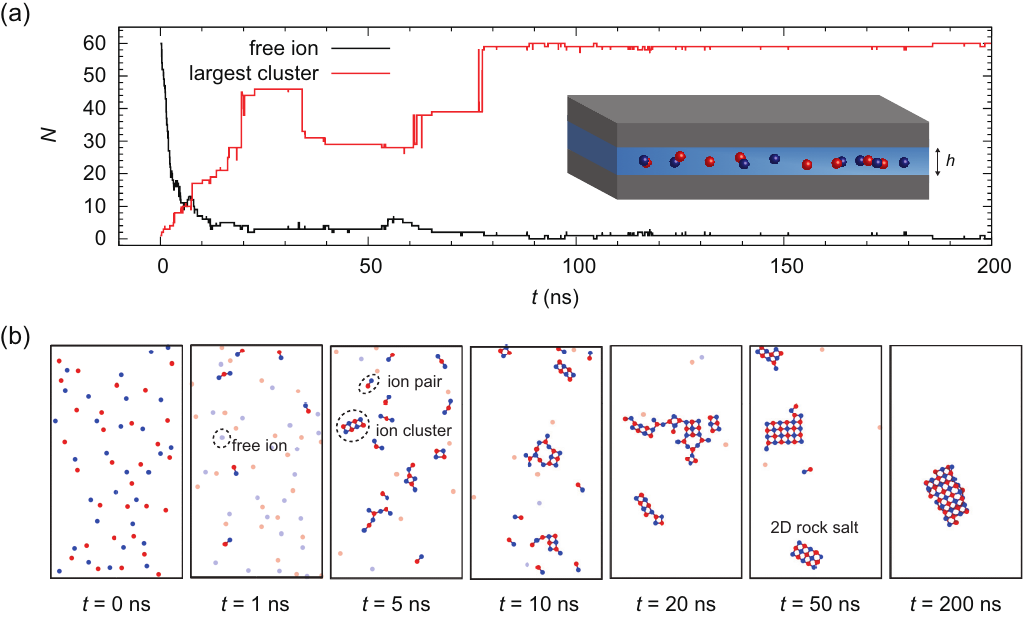}
  \caption{
(a) The evolution of free ions and number of ions within the largest cluster for 30 pairs of Na$^{+}$ and Cl$^{-}$ ions dissolved in monolayer water confined in a 0.8-nm-wide nanochannel.
The inset is the illustration of the nanochannel with the interlayer spacing of $h$.
(b) The simulation snapshots that show the formation of ion pairs or clusters. Cations and anions are colored in red and blue, and free ions are visualized with 30$\%$ transparency.
The ions in the initial system is fully hydrated without ion pairs ($n_{\rm f}=2N$) generated by the modification of cation-anion interaction using the Weeks-Chandler-Andersen (WCA) potential (see details in \textbf{SI}) \cite{wca1971_jcp}. 
The results with different concentrations, electrolytes and simulation details are shown in Figures. S2-S5.
} 
  \label{fig1}
\end{figure*}

We study ionic behaviors in both bulk water and nanoconfined water using a 2D slab model, where $N_{\rm w}$ water molecules and $N$ pairs of ions are confined between rigid walls (inset of Figure \ref{fig1}a) \cite{zhou2020_nanoletter}.
The interlayer spacing ($h$) ranges from $0.65$ nm to $1.35$ nm, corresponding to 1L to multilayer confined water. 
We focus on alkali-metal halide electrolytes like NaCl and KCl, which are common in physiological systems and serve as model ions.
These hydrated ions, with sizes around $0.3$ to $0.6$ nm, are comparable to $h$.
Given that ion hydration within such confined systems is sensitive to ion-wall interaction \cite{qian2023_ic}, we modify their interaction parameters for MD simulation and the distribution of ions within confined water can well match the results of \emph{ab-initio} MD (AIMD) (see details in \textbf{SI} and our recent work \cite{qian2023_ic}).


Building on our previous work on ions in multilayer water ($h \geq 1.0$ nm) \cite{zhou2020_nanoletter}, we now investigate ion pairing in monolayer water using equilibrium MD (EMD).
We start with a fully hydrated state (no pairing, the number of free ions is $n_{\rm f} = 2N$) where the attraction between cations and anions is removed, using the Weeks-Chandler-Andersen (WCA) potential \cite{wca1971_jcp} (see details in \textbf{SI}). 
Figure \ref{fig1} shows that in a system with $N = 30$ Na$^{+}$ and Cl$^{-}$ pairs dissolved in monolayer water ($600$ H$_2$O, $h=0.8$ nm) ions gradually form clusters, eventually creating a large 2D rock salt structure within approximately $100$ ns, despite the low concentration of $c = 2.4$ mol/kg water compared to the saturated solubility of NaCl ($6.15$ mol/kg).
This behavior contrasts with bulk water, where the equilibrium state shows $n_{\rm f}\approx 35$ and $n_{\rm {max}} \approx 5$ (Figure S1).
Herein, we also conducted the simulations for LiCl, KCl and RbCl,  all of which exhibited the same phenomenon (Figure S2).
However, the equilibrium configures of ion clusters and the rate of clustering are ion-specific.
Additional supplementary calculations using different ion parameters and an explicit graphene wall also show consistent results (Figures. S3 and S4). 
These findings confirm that the dynamics of ionic assemblies in 1L water are robust and independent of the chemical nature of the ions and specific simulation details.


\begin{figure*}[tbhp]
  \centering
  \includegraphics[width=0.90\textwidth]{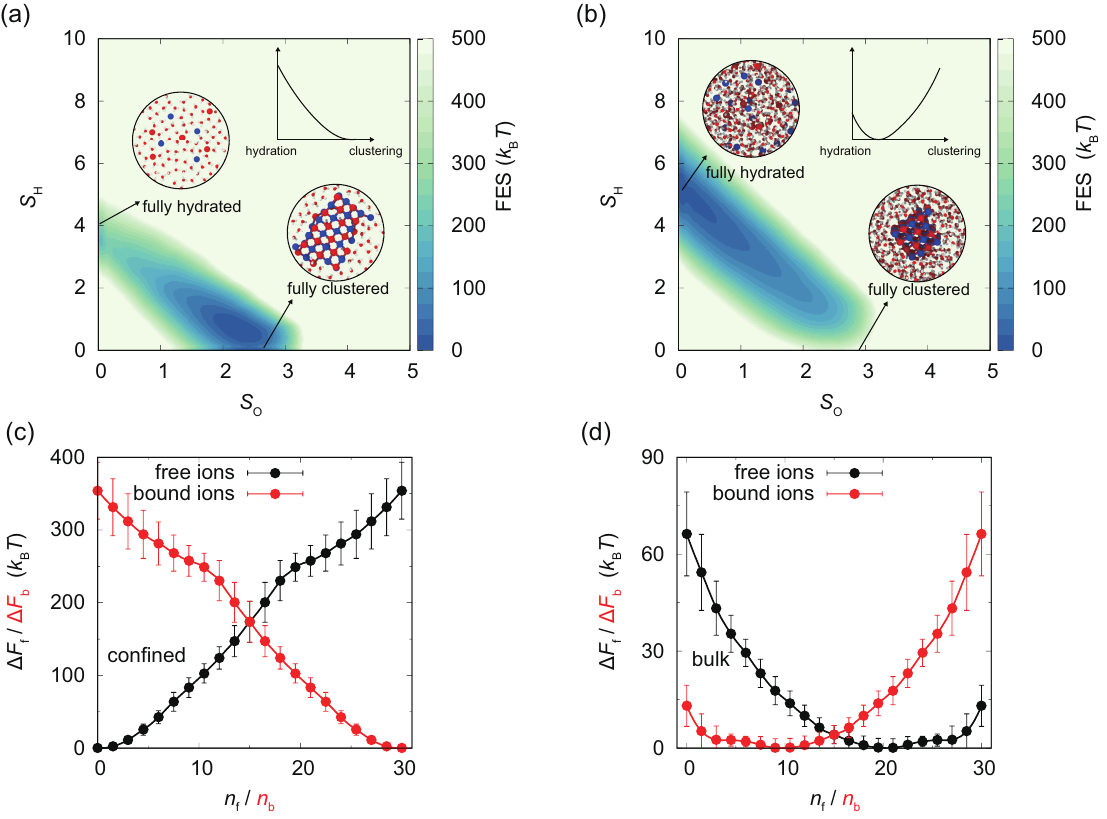}
  \caption{
(a) The 2D free energy surface (FES) as the function of the local order of ions ($S_{\rm O}$) and the degree of ion hydration ($S_{\rm H}$) for 30 pairs NaCl in 1L water with $h=8.5$ {\AA} and (b) bulk water.
The insets with black borders are the snapshots of full hydration state or full clustering states. 
The figures on the upper right corner indicate the free energy level of these two states.
(c) The 1D free energy surface reweighed from the 2D FES as a function of the number of the free ions ($\Delta F_{\rm f}(n_{\rm f})$) or bound ions ($\Delta F_{\rm b}(n_{\rm b})$) for 1L water or (d) bulk water.
The results with different $h$ are shown in Figures. S6-S7.
} 
  \label{fig2}
\end{figure*}

Because the time scale of ion crystallization extends beyond what can be explored in standard simulations and the phase space exploration by EMD is not sufficient enough, the enhanced-sampling MD based on well-tempered Metadynamics (WTMetaD) \cite{welltmp_2008} is performed to explore the energy landscape of ion clustering (see details in \textbf{SI}).
We introduced two order parameters for the crystallization simulations, as inspired by previous work \cite{mp2019_jctc}.
The first one is the degree of ion hydration ($S_{\rm H}$), defined as the number of water molecules within the first hydration shell (1HS), also known as the \emph{hydration number} ($n_{\rm H}$).
The second parameter is the local order of ions ($S_{\rm O}$), defined as the coordination number with counter ions. The exact definitions can be found in the \textbf{SI}.
By conducting long-time WTMetaD simulations (with the total time of more than 3.6 $\mu$s for each system), we can obtain the 2D free energy surface (FES, see Figures. \ref{fig2}a and b).
Notably, we found that the FES configuration for ions in 1L water is reversed compared to that in bulk water.
Specifically, the minimum point for NaCl pairs in 1L water lays near the abscissa, that is $S_{\rm H}\rightarrow 0$, indicating the fully clustered state.
In contrast, for bulk water, the minimum near the ordinate, that is $S_{\rm O}\rightarrow 0$, referring to the fully hydrated state.
The different limiting values for $S_{\rm H}$ when $S_{\rm O}\rightarrow 0$ is because of the different HS structures in 1L water ($<S_{\rm H}>=4.0$) and bulk water ($<S_{\rm H}>=5.7$).   
The markedly different 2D FES for ions in 1L water compared to bulk water directly corroborates the former EMD results.

We can also calculate $n_{\rm f}$ and the number of bound ions ($n_{\rm b}$, which means the ions paired with ions, satisfying $n_{\rm f}+n_{\rm b}=N$) of the system (see details in \textbf{SI}).
Here $n_{\rm{f}}$ and $n_{\rm b}$ only counters for cations or anions.
By reweighing the 2D FES, we can obtain the 1D FES ($\Delta F$) as a function of $n_{\rm f}$ or $n_{\rm b}$ as shown in Figures. \ref{fig2}c and d.
For ions in bulk water, the curve of $\Delta F_{\rm f}(n_{\rm f})$ or $\Delta F_{\rm b}(n_{\rm b})$ is parabolic-like and the minimal point is located at $n_{\rm f}=20.0$ or $n_{\rm b}=10.0$, which means the partial hydration state (or partial clustered state). Moreover, the level of $\Delta F_{\rm f}(n_{\rm f})$ for the full clustered state is larger than the full hydrated state (see the inset of Figure \ref{fig2}b). 
However, the results for ions in 1L water are notably different. 
The FES curve is a monotonic function, with the minimal point located at $n_{\rm f}=0.0$ (or $n_{\rm b}=N$), indicating the full clustered state.
These results explain the EMD findings, where $n_{\rm f}\approx 0$ after long-time simulation, and directly indicate the distinct hydration state in extremely confined water.


\begin{figure*}[tbhp]
  \centering
  \includegraphics[width=0.900\textwidth]{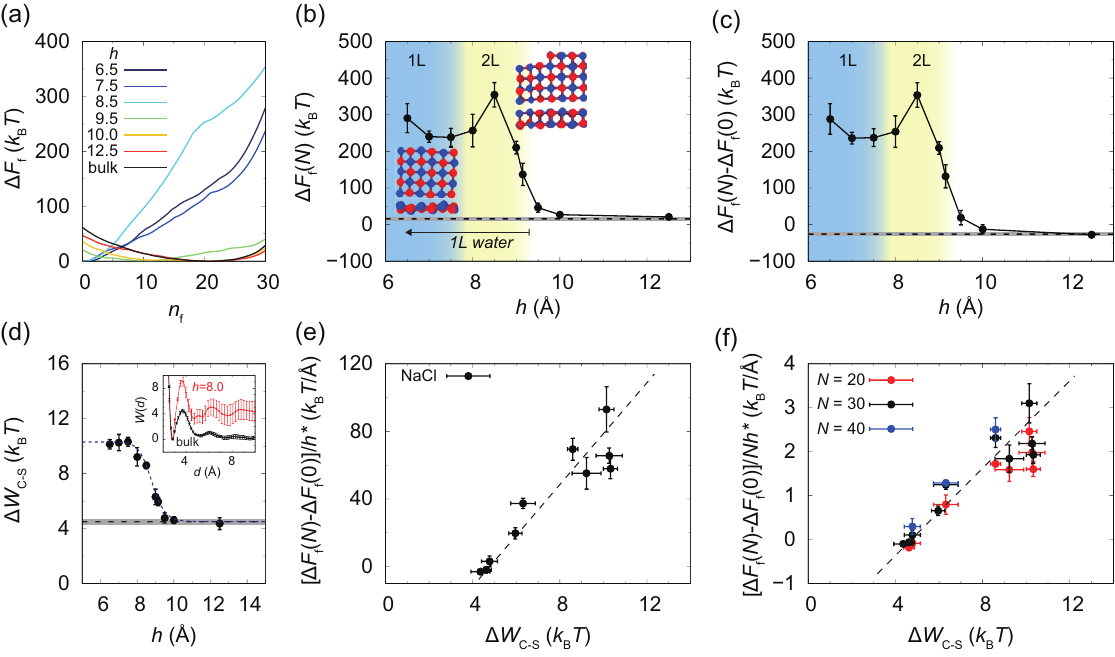}
  \caption{
(a) The 1D free energy surface ($\Delta F_{\rm f}$) as the function of the number of free ions ($n_{\rm f}$) with different $h$ for NaCl.
The results for other $h$ (7.0, 8.0, 9.0, 9.15 {\AA}) are shown in Figure S6.
To clarify, the FES are shown without error bars and the results with errors are shown in Figure S6.  
(b) The results of $\Delta F_{\rm f}(N)$ and (c) $\Delta F_{\rm f}(N)-\Delta F_{\rm f}(0)$ with different $h$.
The insets are the structure of 1L and 2L 2D rock salt.
The background colors indicate the range of 1L and 2L salt.
The black dash lines with the errors by the shadows with 50$\%$ transparency are the result of bulk.
The range of 1L water is also indicated ($h<0.95$ nm).
(d) The barriers from CIPs to SSIPs ($\Delta W_{\rm c\mbox{-}s}$). The insets are the PMFs ($W(d)$) for $h=8.0$ {\AA} and bulk water. 
The black dash lines with the errors by the shadows with 50$\%$ transparency are the result of bulk.
(e) The scaling between $[\Delta F_{\rm f}(N)-\Delta F_{\rm f}(0)]/h^{*}$ and $\Delta W_{\rm c\mbox{-}s}$.
(f) The scaling between $[\Delta F_{\rm f}(N)-\Delta F_{\rm f}(0)]/Nh^{*}$ and $\Delta W_{\rm c\mbox{-}s}$.
} 
  \label{fig3}
\end{figure*}

Because the size of the channel is comparable to, or even reaches, the discrete nature of water molecules and hydrated ions, the structure and dynamics would be influenced by the degree of confinement, or the value of $h$.
Here, we calculate the FES for the system with different values of $h$ (Figures \ref{fig3}a-c).
As shown in Figure \ref{fig3}a, the FESs are significantly affected by $h$.
Specifically, $\Delta F_{\rm f}(n_{\rm f})$ is monotonic with $\Delta F_{\rm f}(0)\approx 0$ when $h<9.5$ {\AA} (corresponding to 1L water), but it suddenly becomes parabolic-like when $h\geq9 .5$ {\AA} (starting to form bilayer, or 2L, water).
The results of 2D FES (as shown in Figure S7) also indicate that the FESs become bulk-like when $h\geq9 .5$ {\AA}.
These results directly demonstrate that the dynamics of ionic assemblies are highly sensitive to confinement.
Herein, we also summarize the result of $\Delta F_{\rm f}(N)$, which refers to the potential to escape from the fully hydrated states, and $\Delta F_{\rm f}(N)-\Delta F_{\rm f}(0)$, which indicates the free energy difference between the fully clustered states and the fully hydrated states (Figures \ref{fig3}b and c).
The result shows $\Delta F_{\rm f}(N)$ (or $\Delta F_{\rm f}(N)-\Delta F_{\rm f}(0)$) is about larger than 200 $k_{\rm B}T$ when $h\lesssim 9.0$ {\AA} but suddenly decreases when $h\gtrsim 9.0$ {\AA} and converges to results of bulk when $h=12.5$ {\AA}.
The high level of these two quantities indicates the ions are ready to drop from the fully hydrated to the clustered state, which is consistent with the results of EMD. 
We also calculate the results at different $c$ or different electrolyte (KCl) and the results are quite similar (Figure S8).
Interestingly, there is a peak when $h=8.5$ {\AA} for both NaCl and KCl.
Observing the structure of the ion clusters, we find that the ions form a 2L 2D rock salt structure (as shown in the insets) when $h\gtrsim 8.0$ {\AA}, but it forms a monolayer when $h\lesssim 7.5$ {\AA}.
The high level when $h=8.5$ {\AA} indicates that the ions are most likely to form clusters at this spacing.
To further confirm the stability of the 2L salt in confined 1L water, supplementary AIMD simulations were carried out (Figure S9). The results show that 2L NaCl is stable within the channel with $h=8.5$ {\AA} at 300 K or even at 390 K.


The formation of ion clusters and the absence of free ions at low $c$ in extremely confined water is unexpected.
To explain and understand this anomalous phenomenon, the evolution of potential energy (or enthalpy, $\Delta H$) is calculated (Figure S10), which is monotonously decreased with $t$ starting from the fully hydrated state.
Because the entropy is reduced in the clustered state, the clustering process is enthalpy-driven.
Here, we also find the ion-ion interaction energy (that is the enthalpy contribution for clustering, $\Delta H_{\rm{i\mbox{-}i}}$) is decreased while the hydration contribution (ion-water and water-water interaction, $\Delta H_{\rm{hy}}=\Delta H_{\rm{i\mbox{-}w}}+\Delta H_{\rm{w\mbox{-}w}}$) is increased (Figure S10a). 
These results mean ion-ion interaction is dominant in 1L water.
While the result of $h=10$ {\AA} is reversed and the hydration effect is dominant (Figure S10b).
We also perform additional simulations as follows.
Firstly, the strength of $\Delta H_{\rm{i\mbox{-}w}}$ is increased to 1.1-fold (by changing the pair interaction between ions and water in MD) or $\Delta H_{\rm{i\mbox{-}i}}$ is reduced to 0.9-fold for NaCl with $h=8.5$ {\AA}. We find the clusters are dissolved in 1L water (Figures S11a-b).
Secondly, the strength of $\Delta H_{\rm{i\mbox{-}w}}$ is reduced to 0.9-fold or increase the $\Delta H_{\rm{i\mbox{-}i}}$ to 1.2-fold for $h=1.0$ nm. We find the dissolved NaCl can assemble to cluster (Figures. S11c-d).
From the above analysis, the state of the ions (free ions or clusters) is determined by the competition between ion hydration and ion clustering (ion-ion interaction). In 1L water, ion clustering is dominant, whereas in thicker water layers, ion hydration prevails.

Here, we also calculate the potential of means force (PMF), $W(d)$, for the single ion pair as a function of cation-anion distance ($d$) in the channel with different $h$ (the inset of Figures \ref{fig3}d and S12).
It shows apparent valleys at $d \approx 3$ and $\approx 5$ {\AA} for NaCl, corresponding to the association of contact ion pairs (CIPs) and solvent-shared/separated ion pairs (SSIPs).
We find the barriers from CIPs to SSIPs ($\Delta W_{\rm c\mbox{-}s}$) is about 10$k_{B}T$ when $h\lesssim 8.0$ {\AA} and suddenly decreased before converge to bulk value at $h=12.5$ {\AA}.
The deep valleys for CIPs make it difficult for ion pairs to de-pair once they have paired.
In fact, the barrier of 5$k_{B}T$ is already high enough to result in good rejection rates, greater than $99\%$ (such that 1-$\exp(-\Delta W_{\rm c\mbox{-}s}/k_{\rm B}T)>99 \%$). 
The association constants $K_{\rm a}$ can be calculated from $W(d)$ \cite{justice1976_jsc}, that is 
$K_{\rm a}=4\pi R \int_{0}^{r_{\rm c}}\exp(-W(r)/k_{\rm B}T)r^{2}dr$, 
where $R={\rm {N_{A}}}\times 10^{-24}$ (N$_{\rm A}$ is the Avogadro number), 
$r_{\rm c}$ is the cation-anion distance of criterion dividing the CIP and SSIP.
Thus the value of $\Delta W_{\rm c\mbox{-}s}$ can measure the stability of clustering.
The deep valleys of CIPs result in the extremely large value of $K_{\rm a}$, which also explains the spontaneous formation of ion clusters in 1L water.

However, the result of PMFs cannot explain the anomalous increase of $\Delta F_{\rm f}(N)$ (or $\Delta F_{\rm f}(N)-\Delta F_{\rm f}(0)$) at $h\approx 8.5$ {\AA}.
These results predict that $\Delta F_{\rm f}(N)$ should decrease with the increase of $h$, given the change of $\Delta W_{\rm c\mbox{-}s}(h)$.
Thus, the clustering mechanisms cannot be simply attributed to the PMFs of single pairs or two-body interactions.
Instead, they likely involve collective or many-body behaviors of the system. Moreover, there may be a positive contribution to ion clustering.
In this work, we explain this phenomenon using classical nucleation theory (CNT).
The free energy cost of the nucleus for nucleation in bulk water can be expressed as, 
$\Delta G \sim -4/3{\pi}r^{3}\Delta \mu+4{\pi}r^{2}\sigma$,
where $r$ is the size of nucleus, $\Delta\mu$ is the volume contribution and $\sigma$ is the surface contribution.
One point should be noted that $\Delta F$ calculated in this work is not equivalent to the $\Delta G$, which can only be determined from the constant chemical potential molecular dynamics (C$\mu$MD) \cite{mp2019_jctc}. 
However, CNT can help clarify the underlying mechanisms.
In strong planar confinement with 1L water,
$\Delta G \sim -4{\pi}r^{2}h^{*}\Delta \mu+2{\pi}rh^{*}\sigma \sim h^{*}$, 
where $h^{*}$ is the accessible width for ion clustering.
The contribution of ion-wall interaction is ignored because its values are much smaller than the other two terms and remains constant during clustering (Figure S6). 
Here $h^{*}$ is the effective width of channel, defined as $h^{*}=h-3.4$ {\AA}, 3.4 {\AA} is the thickness of graphene channel. 
Therefore, $h$ is another factor and it contributes positively.
Combining this with the results from the previous paragraph, we can get the scaling law,
$\Delta F(N) \sim \Delta W_{\rm c\mbox{-}s} \times h^{*}, \Delta F(N)/h^{*} \sim \Delta W_{\rm c\mbox{-}s}$. 
This scaling law is well confirmed by our simulation results (Figures \ref{fig3}e and S12). 
It is easy to know $\Delta F(N) \sim N$. Then $\Delta F(N)/h^{*}/N \sim \Delta W_{\rm c\mbox{-}s}$. This concentration-independence scaling is also confirmed by our results (Figure \ref{fig3}f).
These results imply the existence of the optimal $h$ for ion clustering, which we find to be 8.5 {\AA} for both NaCl and KCl, forming a bilayer 2D rock salt structure.


\begin{figure*}[tbhp]
  \centering
  \includegraphics[width=0.900\textwidth]{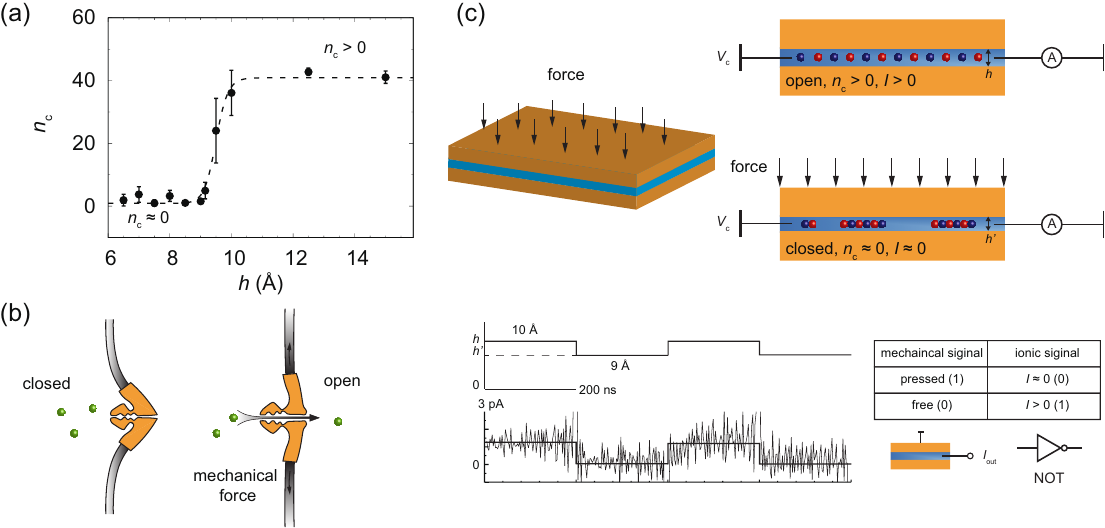}
  \caption{
(a) The number of effective charge carrier ($n_{\rm c}$) as a function of $h$. The system contains 30 pairs of NaCl and 600 water molecules.
(b) The illustration of biological mechanosensitive channels that the passage of ions is modulated through changing their spatial configurations by the force from lipids \cite{epand2018}.
(c) The conceptual mechanosensitive ion channel, where $V_{\rm c}$ is the translational voltage.
The pairing effect is tuned by nanoconfinement (that is the value of $h$) that $n_{\rm c}\approx 0 $ when $h\lesssim 9.0$ {\AA} but become a finite value when $h\gtrsim 9.5$ {\AA}.
Here the $h$ changes between 10.0 {\AA} and 9.0 {\AA} that $I$ changes between $\approx1.25$ nA and 0 under the driven electric field of 0.005 V/{\AA}.
From the reverse response ($0 \rightleftharpoons 1$) of ionic signal ($I$) to mechanical signal, the ionic mechanically \texttt{NOT} logic gate can be constructed. 
} 
  \label{fig4}
\end{figure*}

The confinement-sensitive ion clustering discovered in this work indicates that the charge carriers in a confined hydrated ionic system can be effectively controlled by the degree of confinement.
The number of effective charge carrier $n_{\rm c}$ can be estimated as 
$ n_{\rm c}=2 \times \int_{0}^{N}n_{\rm f} \times \exp(-\Delta F_{\rm f}(n_{\rm f})/k_{\rm B}T)/Z{\rm{d}}n_{\rm f}$,
where $Z=\int_{0}^{N} \exp(-\Delta F_{\rm f}(n_{\rm f})/k_{\rm B}T) {\rm{d}}n_{\rm f} $ is the partial function.
The contribution of bound ions is ignored because the effective charge of bound ions is zero (if the number of cation and anion is equal) or the mobility of bound ions with large size is much smaller than free ions.  
The results in Figure \ref{fig4}a show the $n_{\rm c}$ is sensitive to $h$ and exhibits a switch-like behavior that $n_{\rm c}\approx 0 $ when $h\lesssim 9.0$ {\AA} but become a finite value when $h\gtrsim 9.5$ {\AA}. This indicates that the open and closed state can be effectively controlled by $h$.
The piezo channels in biological membranes control the passage of ions by altering their spatial configurations as shown in Figure \ref{fig4}b.
Following the principles discovered in this work, we design a mechanosensitive ion channel as shown in Figure \ref{fig4}c. This design is also confirmed by MD simulations (see \textbf{Supplementary Video}). 
While a translational voltage ($V_c$) is applied along the channel, free cations and anions will be driven to move along opposite directions and generate ionic currents ($I>0$).
Under pressure, when $h$ decreases to 9.0 {\AA}, ions could be cluster with $n_{\rm c}\approx 0 $, resulting in $I\approx 0$.
Consequently, this setup rectifies the ionic current under the control of external force.
The open and closed states can be precisely controlled by the external mechanical stimuli by simply changing the $h$ by 1 {\AA} (if changes between 10.0 {\AA} and 9.0 {\AA}), achieving the theoretical on-off ratio ($R$, equal to the ratio of $n_{\rm c}$) on the order of 100.
The value is much larger than the results if only considering the change of pore geometry ($R=(10-3.4)/(9.0-3.4)=1.18$).
From the reverse response ($0 \rightleftharpoons 1$) of $I$ to pressure signal, the \emph{ionic mechanically logic gate} (here is \texttt{NOT}) can be constructed. 
The other two simplest logical gates, \texttt{NAND} and \texttt{NOR}, can also be programmed through the parallel and series circuits of the elementary devices, respectively \cite{zhou2020_nanoletter}.
Our idea would pave the way for the development of more complex iontronic mechanotransduction devices on nanofluidic chips using advanced circuitry, such as nanofluidic mechanical computation.




While ion clustering in confined water is unexpected, some potential experimental evidence exists.
Experimental work finds the salt (for example, NaCl) contaminants precipitate as nanocrystals in the dried-out graphene liquid cells \cite{zhou2015_nature}.
Analysis of the diffraction data shows that these NaCl nanoplatelets display a square lattice with a spacing of approximately 2.8 Å, which matches the equilibrium distance of NaCl ion clusters according to our MD simulations \cite{zhou2020_nanoletter}.
Recently, L. Wang \emph{et al.} also find the crystallization of rock salt NaCl in situ graphene liquid cell transmission electron microscopy (TEM) \cite{wang2021_prl}.
In addition, two-dimensional NaCl crystals is discovered on graphene oxide membranes in dilute solution at ambient conditions \cite{fhp2018_natchem}.
Moreover, unexpectedly high salt accumulation inside carbon nanotubes in dilute salt solutions is discovered by high-angle annular dark field scanning transmission electron microscopy (HAADF-STEM) image \cite{fhp2018_prl}.
Additionally, ion clusters or chains have been observed near the surfaces of mica and Au, as directly measured by atomic force microscopy (AFM) \cite{tian2024_natnano,ricci2014_nc}.
These microscopy experiments imply the presence of ion clustering in confined water.
K. Gopinadhan \emph{et al.} find ions can be completely sterically excluded through confined 1L water, while the entrance barrier for Na$^+$ and K$^+$ is $\approx 0.25$ eV according to the results of classical MD \cite{zhou2020_jpcc,abraham2017_natnanotech} (the true barriers would be much lower than 0.25 eV if consider cation-$\pi$, which is on the order of $\gtrsim 1.0$ eV \cite{zhou2020_prr}) that it is not particularly high and is similar as the activation energy of proton or diffusion barriers of ions in metal organic framework \cite{loewenstein1962_jacs,miner2019_jacs}.
Based on our results, we attribute the impediment for ions to the formation of large ion clusters with low diffusion rates or with near zero effective charge and low mobility.
However, these experimental results are not direct evidence. 
Advanced electron microscopy tools for imaging nanoscale structures and transport, or ultra-fast spectroscopy technologies, are needed to further investigate the structures and transport of hydrated ions under extreme confinement.

\section*{Summary}
In summary, we find the spontaneous formation of ion clusters in 1L water while not for thicker water layers by comprehensive enhanced-sampling MD simulations.
Using free energy calculation, the hydration state of ions as free ions or clusters is determined by the competition between ion hydration and ion-ion interaction that the latter is dominant in 1L water while not in thicker water layers.
We also find that ion clustering is confinement-sensitive, which means the charge carrier of the system can be well regulated by the confinement.
Following these principles, we designed mechanotransduction ion channel that the external mechanical stimuli can well control the open or closed states.
These results establish foundational principles for the mechanosensing of piezo-channels and electromechanical coupling in biological systems, extending beyond the perspective of conformational changes (i.e., simply changing their spatial configurations).
Furthermore, the utilization of water and ions in nanofluidic systems, akin to those found in biological systems, hints at the potential for interfacing artificial channels with biological systems. 
Although our findings are based on computational investigations, we believe their implementation is feasible using current experimental techniques involving van der Waals assembly and well-aligned 2D materials membranes\cite{bocquet2019nature,geim2018_science,geim2017_science,
bocquet2023_science,chenqf2024_science}.


\section*{Acknowledgements}

The authors acknowledge the financial support of the National Natural Science Foundation of China (12572126). 
The work was carried out at the National Supercomputer Center in Tianjin, and the calculations were performed on TianHe-1(A).

\section*{Author contributions}

KZ conceived the project, carried out the molecular dynamics simulations and theoretical analysis, wrote the paper. 


\section*{Data availability}
The data within this paper and other findings of this study are available from the corresponding authors upon reasonable request.

\section*{Code availability}
The codes and input scripts used within this work are available from the corresponding authors upon reasonable request.

\section*{Conflicts of interest}
There are no conflicts to declare.





\providecommand{\latin}[1]{#1}
\makeatletter
\providecommand{\doi}
  {\begingroup\let\do\@makeother\dospecials
  \catcode`\{=1 \catcode`\}=2 \doi@aux}
\providecommand{\doi@aux}[1]{\endgroup\texttt{#1}}
\makeatother
\providecommand*\mcitethebibliography{\thebibliography}
\csname @ifundefined\endcsname{endmcitethebibliography}
  {\let\endmcitethebibliography\endthebibliography}{}

\end{document}